\documentclass[superscriptaddress,amsfonts,amssymb,amsmath,twocolumn,floatfix]{revtex4}

\usepackage{soul}
\usepackage{ulem}
\usepackage{xcolor}
\usepackage{amsfonts} 
\usepackage[makeroom]{cancel}
\usepackage{txfonts}

\newcommand{\ou}{
  \mathrel{
    \vcenter{\offinterlineskip
      \ialign{##\cr$\prec$\cr\noalign{\kern-1.5pt}$\succ$\cr}
    }
  }
}

\usepackage[USenglish]{babel}
\usepackage{epsfig}
\usepackage{graphics}
\usepackage{graphicx} 
\usepackage{amsmath}
\usepackage{color}
\usepackage{comment}
\usepackage{slashed}
\usepackage{siunitx}
\usepackage{hyperref}

\usepackage{times}
\usepackage{tikz}\usetikzlibrary[arrows]\usetikzlibrary{decorations.markings}
\usepackage{mathrsfs}
\usepackage{url}
\usepackage{comment}
\usepackage{appendix}
\usepackage{braket}

\hypersetup{
    colorlinks=true,
    linkcolor=blue,
    citecolor=blue,
    filecolor=magenta,
    urlcolor=cyan,
    pdfpagemode=FullScreen,
    }

\date{\today}

\begin{document}

\title{Larmor radiation as a witness  to the Unruh effect }


\author{Atsushi Higuchi}
\email{atsushi.higuchi@york.ac.uk}
\affiliation{Department of Mathematics, University of York, Heslington, York YO105DD, United Kingdom}

\author{George E. A. Matsas}
\affiliation{Instituto de Física Teórica, 
Universidade Estadual Paulista, Rua Dr. Bento Teobaldo Ferraz, 271, 01140-070, São Paulo, São Paulo, Brazil}

\author{Daniel A. T. Vanzella}
\affiliation{Instituto de F\'\i sica de S\~ao Carlos, Universidade de S\~ao Paulo, \\
Caixa Postal 369, 13560-970, S\~ao Carlos, SP, Brazil}
\affiliation{Institute for Quantum Optics and Quantum Information (IQOQI), 
Austrian Academy of Sciences, Boltzmanngasse 3, A-1090 Vienna, 
Austria\footnote{On a sabbatical leave.}}

\author{Robert Bingham}
\affiliation{Rutherford Appleton Laboratory, Chilton, Didcot, Oxon OX11OQX, UK} \affiliation{Department of Physics, University of Strathclyde, Glasgow G40NG, UK}

\author{Jo\~ao P. B. Brito}
\affiliation{Programa de P\'os-Gradua\c c\~ao em F\'{\i}sica, Universidade Federal do Par\'a, 66075-110, Bel\'em, Par\'a, Brazil}

\author{Lu\'{\i}s C. B. Crispino}
\affiliation{Programa de P\'os-Gradua\c c\~ao em F\'{\i}sica, Universidade Federal do Par\'a, 66075-110, Bel\'em, Par\'a, Brazil}

\author{Gianluca Gregori}
\affiliation{Department of Physics, University of Oxford, Parks Road, Oxford OX13PU, UK}

\author{Georgios Vacalis}
\affiliation{Department of Physics, University of Oxford, Parks Road, Oxford OX13PU, UK}


\pacs{}
\begin{abstract}
We discuss the emission of radiation from general sources in quantum scalar, electromagnetic and gravitational fields using 
the Rindler coordinate frame, which is suitable for uniformly accelerated observers, in the Minkowski vacuum.  In particular, we point out that, to recover, from the point of view of uniformly accelerated observers in the interaction picture, the usual Larmor radiation, which is independent of the choice of the vacuum state, it is necessary to incorporate the Unruh effect assuming the Minkowski vacuum.  Thus, the observation of classical Larmor radiation in the Minkowski vacuum could be seen as vindicating the Unruh effect in the sense that it is not correctly recovered in the uniformly accelerated frame unless
the Unruh effect is taken into account.
\end{abstract}

\maketitle

     \section{Introduction}
     \label{introduction}

Quantum mechanics (QM) and special relativity (SR),  each in their own way, 
have profoundly changed our understanding of Nature.  While classic determinism had to 
give way to the inherent uncertainties of quantum superpositions, 
space and time 
were eventually understood as just  parts  of a more fundamental stage, the spacetime,  
on which 
mass-energy equivalence opened the way for ``{matter}'' to be no longer conserved.  
In view of these shifts in the paradigm separately promoted by QM and SR, it should not 
come as a surprise that their  combination, into the successful framework of quantum field 
theory (QFT),  would lead to novel effects which 
defy our intuition. The relative (hence,  non-fundamental) nature of the quantum particle 
concept is possibly one of the most underappreciated revelations of QFT.  It probably finds 
its utmost realization in the Unruh effect~\cite{unruh_1976},  
according to which the usual vacuum state of a QFT in 
Minkowski spacetime---i.e.\ the state that inertial observers describe as the absence 
of real particles---is ``seen'' as a thermal bath of particles by uniformly accelerated 
observers. 
This effect is closely related to the Hawking effect, which leads to the evaporation
of black holes~\cite{hawking_1974,hawking_1976}.

Despite the  rigorous derivations of the Unruh effect in QFT using 
different approaches (see Refs.~\cite{bisognano_1975, bisognano_1976, sewell_1982, fulling_2004}), its reality (or consequences) are
frequently put into question (see, e.g., Ref.~\cite{belinskii_1997,narozhny_2001,ford_2006,gelfer_2015,cruz_2016,connel_2020,popruzhenko_2023}) --- not rarely due to misconceptions about its 
precise meaning.
These debates have motivated several proposals for observing 
signatures of the acceleration radiation in the laboratory (see, e.g., 
Refs.~\cite{bell_1983, vanzella_2001, suzuki_2003, cozzella_2018, chen_1999,
schutzhold_2006, cozzella_2017, lima_2019, lynch_2021, leonhardt_2018, barros_2020} and references therein).
In considering these proposals,  it is important to keep in mind that the 
Unruh effect is {\it not} an additional ingredient which is assumed on  top of 
QFT in the inertial frame;  this is an effect that must be taken into account 
when interpreting standard inertial-frame QFT results from the perspective of 
uniformly accelerated observers.  In other words,  without the Unruh thermal 
bath in the uniformly accelerated frame, uniformly accelerated observers
would not be able to explain the phenomena that inertial observers in the usual vacuum
successfully describe with the standard QFT. 
The Unruh effect in QFT is analogous to
{\it inertial} forces,  also known as {\it fictitious forces},  in Newtonian mechanics:  
one needs to take the inertial forces into account  in the non-inertial frame in order to correctly describe the motion 
of particles which inertial observers successfully describe using Newton's laws 
without these ``extra'' forces.  Similarly, one needs to take the Unruh effect into account to use QFT in a uniformly accelerated frame in order
to correctly describe the results of QFT in an intertial frame.
This analogy clarifies what constitutes an 
``observation'' of the Unruh effect in {\it any} inertial-lab-based experiment:
it means measuring an effect in the inertial frame which uniformly accelerated 
observers can only account for by using the Unruh thermal bath.
 
 The main purpose of this paper is to explain further the fact 
that usual {\it classical} radiation (as measured 
in inertial frames), such as 
Larmor radiation~\cite{larmor_1897,jackson_1998}, 
can be considered as such an observation of the Unruh effect.  This has 
already been pointed out in the literature with particular sources and 
worldlines~\cite{higuchi_1992R, higuchi_1992, ren_1994, landulfo_2019, 
portales_oliva_2022, vacalis_2023, portales_oliva_2024, brito_2024}; here, 
we present a general proof valid not only for the electromagnetic field with general charge distributions  but also for the 
scalar and graviton fields.

Building on the observation of Unruh and Wald that the emission of a quantum in the inertial 
frame corresponds to either the emission or absorption of a quantum to or from the Unruh 
thermal bath in the uniformly accelerated frame~\cite{unruh-wald_1984}, we show 
here
that, at first order in perturbation theory,
the {\it interaction probability}
(i.e.\ the sum of emission and absorption probabilities) 
of a classical source
according to uniformly accelerated observers, in the presence of the Unruh thermal bath, 
gives exactly the emission probability needed to reproduce classical radiation
from the inertial perspective. 
The quantum nature of the Unruh effect is reflected in the fact that this agreement 
is only possible by assuming that radiation is made of {\it quanta}, whose
energy $E$  and frequency $\nu$  are related by Planck's  formula, $E = h\nu$,  
with $h$ being Planck's constant.

The rest of the paper  is organized as follows.  In Sec.~\ref{sec:sum},  we expand the 
massless Klein-Gordon scalar field using the Unruh modes and their complex conjugates, which are eigenfunctions of the boost Killing vector field and are  
positive- and negative-frequency solutions with respect to inertial time translations---and, 
as such, can be used to define the vacuum state, representing absence of particles 
according to inertial observers. 
In Sec.~\ref{sec:sum_rc},  we relate the Unruh modes with  the Rindler modes,  
which are eigenfunctions of the boost Killing vector field with support in a Rindler wedge%
---hence, related to quantization in the uniformly accelerated frame.  
Then,  in Sec.~\ref{sec:clas_rad}, we calculate the following two quantities: (i)~the  probability 
$P_{\rm em}^{(\rm M)}$ for a classical source to emit a quantum of the field, from the 
inertial-frame perspective, and (ii) the 
probability $P_{\rm int}^{(\rm R)}$ for the same classical source to absorb or emit
a quantum from or to the Unruh thermal bath, from the uniformly accelerated perspective. 
These two quantities turn out to be exactly the same, provided the source is completely 
contained in the Rindler wedge (so that it can be fully described in the uniformly 
accelerated frame). 
Sec.~\ref{sec:em_gr} is devoted to extending the  result
in Sec.~\ref{sec:clas_rad}
to the electromagnetic and graviton fields.  Finally,  in Sec.~\ref{sec:discuss} we 
present our final considerations and discussion. In Appendix~\ref{app-A} 
we confirm for the scalar case that the classical radiation formula
is found in the Heisenberg picture in
any vacuum state, and
in Appendix~\ref{app-B} we show, also for the scalar case, how the classical radiation
formula is reproduced in the
Fulling vacuum state as well in the interaction picture.  We  adopt 
the metric signature $(+,-,-,-)$ and natural units, in which
$\hbar = c  = 1$.

     \section{The Unruh modes as the boost eigenfunctions}
     \label{sec:sum}

We begin by considering  the Unruh effect  
for the free massless scalar field.
This quantum field is naturally expanded in terms of the momentum eigenstates 
as~\cite{peskin-schroeder}
\begin{align}
    \widehat{\phi}(x) = \int \frac{d^3\mathbf{k}}{(2\pi)^3 2k}
    \left[ f^{\mathbf{k}}(x)\,\hat{a}_\mathbf{k} 
    + \overline{f^{\mathbf{k}}(x)}\,\hat{a}^\dagger_{\mathbf{k}} \right]\,,
    \label{eq:phi-expansion-in-k}
\end{align}
with $x=(t,\mathbf{x}) = (t,z,\mathbf{x}_\perp)$,
$\mathbf{k} = (k_z,\mathbf{k}_\perp)$, and $k=\|\mathbf{k}\|$, 
where
\begin{align}
    f^{\mathbf{k}}(x) = e^{-ikt + i\mathbf{k}\cdot \mathbf{x}}
    =e^{-ikt + i k_z z+ i\mathbf{k}_\perp\cdot \mathbf{x}_\perp} \,.
\end{align}
The annihilation and creation operators, $\hat{a}_{\mathbf{k}}$ 
and $\hat{a}^\dagger_{\mathbf{k}}$, satisfy
\begin{align}
    [\hat{a}_{\mathbf{k}},\hat{a}^\dagger_{\mathbf{k}'}] 
    = 
    (2\pi)^32k\,\delta^{(3)}(\mathbf{k}-\mathbf{k}')\,,  
    \label{eq:the-commutator}
\end{align}
with all other commutators among them vanishing.  The Minkowski vacuum, 
$\ket{0_\mathrm{M}}$, is defined by 
\begin{align}
    \hat{a}_\mathbf{k}\ket{0_\mathrm{M}} = 0\,\ \textrm{for all}\ \mathbf{k}\,. 
\end{align}

The Unruh modes, 
introduced in Unruh's original paper~\cite{unruh_1976}, 
can be characterized 
as the superpositions of $f^{\mathbf{k}}(x)$ that are eigenfunctions of the boost 
transformation in the $z$-direction.
 To see this, 
we first define the rapidity in the $z$-direction by
\begin{align}
    \vartheta & = \frac{1}{2a}\ln \frac{k+k_z}{k-k_z}\,,
\end{align}
where $a\,(>0)$ and $\vartheta$ have the dimensions of acceleration and time, 
respectively. 
 Then,
\begin{align}
    k & = k_\perp \cosh a\vartheta\,, \label{eq:k0-in-rapidity}\\
    k_z & = k_\perp\sinh a\vartheta\,, \label{eq:kz-in-rapidity}
\end{align}
where 
$k_\perp = \|\mathbf{k}_\perp\|$.
The commutator~\eqref{eq:the-commutator} can be written as
\begin{align}
     [ \hat{a}_{\mathbf{k}},\hat{a}^\dagger_{\mathbf{k}'}] 
     = 16\pi^3 a^{-1}\delta(\vartheta-\vartheta')
     \delta^{(2)}(\mathbf{k}_\perp-\mathbf{k}'_\perp)\,,  \label{eq:a-commutator}
\end{align}
where $\vartheta'$ is the rapidity for $\mathbf{k}'$, and the measure for the integration 
over $\mathbf{k}$ can be written as
\begin{align}
    \frac{d^3\mathbf{k}}{(2\pi)^32k} 
    = 
    \frac{a}{16\pi^3}d\vartheta\,d^2\mathbf{k}_\perp\,. \label{eq:measure-in-rapidity}
\end{align}

Now, we define the Unruh modes by~\cite{crispino_2008,Higuchi:2017gcd}
\begin{align}
    u^{(\omega,\mathbf{k}_\perp)}(x) & = a\int_{-\infty}^\infty 
    e^{-i\omega\vartheta}f^{\mathbf{k}}(x)\,
    d\vartheta\,, \label{eq:def-of-Unruh}
\end{align}
which can be inverted as
\begin{align}
    f^{\mathbf{k}}(x) = \frac{1}{2\pi a} \int_{-\infty}^\infty 
    e^{i\omega\vartheta}u^{(\omega,\mathbf{k}_\perp)}(x)\,d\omega\,.
    \label{eq:def-of-Unruh-inverted}
\end{align}
(See Ref.~\cite{Ueda:2021nln} for a similar formula for the Unruh modes for the spinor field.)
Under the boost transformation parametrized by $\beta$,
\begin{align}
    (t,z) & \mapsto (t\cosh a\beta - z\sinh a\beta, -t\sinh a\beta + z\cosh a\beta)\,,
    \label{eq:boost-transformation}
\end{align}
we have $f^{\mathbf{k}}(x)\mapsto f^{\mathbf{k}'}(x)$, where 
$$
\mathbf{k}'_\perp = \mathbf{k}_\perp 
\quad
{\rm and} 
\quad
k'_z = k_\perp \sinh[a (\vartheta + \beta) ].
$$
This transformation can be undone in the integral in Eq.~\eqref{eq:def-of-Unruh} 
by letting $\vartheta \mapsto \vartheta - \beta$. Thus, under the boost 
transformation~\eqref{eq:boost-transformation} the Unruh modes transform as
$u^{(\omega,\mathbf{k}_\perp)}(x)\mapsto e^{i\omega\beta}u^{(\omega,\mathbf{k}_\perp)}(x)$.  
That is, the Unurh modes are indeed eigenfunctions of the boost transformation 
in the $z$-direction.

It is clear from Eqs.~\eqref{eq:def-of-Unruh} and \eqref{eq:def-of-Unruh-inverted} 
that the Unruh modes form a (generalized) basis for the space of positive-frequency
solutions to the Klein-Gordon equation.  Hence, one can expand the quantum 
field $\widehat{\phi}(x)$ in terms of the Unruh modes:
\begin{equation}
    \widehat{\phi}(x) 
    = \int \frac{d^2\mathbf{k}_\perp}{16\pi^3 a} \int_{-\infty}^\infty d\omega 
    \left[ u^{(\omega,\mathbf{k}_\perp)}
    (x)\,\hat{b}_{(\omega,\mathbf{k}_\perp)}
    + 
    \overline{u^{(\omega,\mathbf{k}_\perp)}(x)}\,\hat{b}^\dagger_{(\omega,\mathbf{k}_\perp)}\right]. 
    \label{eq:phi-expansion-in-omega}
\end{equation}
Substituting Eq.~\eqref{eq:def-of-Unruh-inverted} into Eq.~\eqref{eq:phi-expansion-in-k} 
with Eq.~\eqref{eq:measure-in-rapidity}
and identifying the coefficients of $u^{(\omega,\mathbf{k}_\perp)}(x)$ as
$\hat{b}_{(\omega,\mathbf{k}_\perp)}$ in Eq.~\eqref{eq:phi-expansion-in-omega}, we find
\begin{align}
    \hat{b}_{(\omega,\mathbf{k}_\perp)} 
    = 
    \frac{a}{2\pi}\int_{-\infty}^\infty e^{i\omega\vartheta} \hat{a}_{\mathbf{k}}\,d\vartheta\,.
\end{align}
Then, we find from Eq.~\eqref{eq:a-commutator}
\begin{align}
    [ \hat{b}_{(\omega,\mathbf{k}_\perp)}, \hat{b}^{\dagger}_{(\omega',\mathbf{k}'_\perp)}]
    = 8\pi^2 a\,\delta(\omega-\omega')\delta^{(2)}(\mathbf{k}_\perp-\mathbf{k}'_\perp)\,. 
    \label{eq:b-commutator}
\end{align}
Since the operators $\hat{b}_{(\omega,\mathbf{k}_\perp)}$ are superpositions 
of $\hat{a}_\mathbf{k}$, the Minkowski vacuum state is annihilated by them:

\begin{align}
    \hat{b}_{(\omega,\mathbf{k}_\perp)}\ket{0_\mathrm{M}} 
    = 
    0\,\ \textrm{for all}\ \omega\ \textrm{and}\ \mathbf{k}_\perp\,.
\end{align}

     \section{The Unruh modes in Rindler coordinates}
     \label{sec:sum_rc}

In this section we show that the 
definition~\eqref{eq:def-of-Unruh} of the 
Unruh modes agrees with the standard definition as linear combinations of 
the Rindler modes~\cite{fulling_1973}. Then we review some aspects of the 
Unruh effect. First we define the Rindler coordinates and Rindler wedges.
Right Rindler coordinates, $\tau$ and $\xi$, are defined by the relations,
\begin{align}
    t & = a^{-1}e^{a\xi}\sinh a\tau\,, \label{eq:t-in-Rindler}\\
    z & = a^{-1}e^{a\xi}\cosh a\tau\,. \label{eq:z-in-Rindler}
\end{align}
The constant $a$ is the proper acceleration of the worldline defined by
$\xi=0$ and $\mathbf{x}_\perp=\mathrm{constant}$.  These coordinates, 
together with $\mathbf{x}_\perp$, cover the right Rindler wedge defined by 
$z > |t|$.  Left Rindler coordinates, $\tilde{\tau}$ and $\tilde{\xi}$, 
are defined by
\begin{align}
    t & = a^{-1}e^{a\tilde{\xi}}\sinh a \tilde{\tau}\,,\\
    z & = - a^{-1}e^{a\tilde{\xi}}\cosh a \tilde{\tau}\,. \label{eq:z-in-Rindler2}
\end{align}
These coordinates, together with $\mathbf{x}_\perp$, cover the left Rindler 
wedge defined by $z < -|t|$.

The Unruh modes~\eqref{eq:def-of-Unruh} can be expressed in Rindler coordinates given by
Eqs.~\eqref{eq:t-in-Rindler}--\eqref{eq:z-in-Rindler2},
using  Eqs.~\eqref{eq:k0-in-rapidity} and \eqref{eq:kz-in-rapidity}, as
\begin{align}
    u^{(\omega,\mathbf{k}_\perp)}(x) & = ae^{-i\omega\tau}\int_{-\infty}^\infty
    e^{-i\omega s + i(k_\perp/a)\exp({a\xi})\sinh as 
    + i\mathbf{k}_\perp\cdot\mathbf{x}_\perp}
    \,ds\,,
\end{align}
in the right Rindler wedge, and 
\begin{align}
    u^{(\omega,\mathbf{k}_\perp)}(x) 
    & = ae^{i\omega\tilde{\tau}}\int_{-\infty}^\infty
    e^{-i\omega s - i(k_\perp/a)\exp ({a\tilde{\xi}}) \sinh as 
    + i\mathbf{k}_\perp\cdot\mathbf{x}_\perp}
    \,ds\,, 
    \label{eq:u-left-Rindler}
\end{align}
in the left Rindler wedge. Using the formula~\cite[Eq.~10.32.7]{NIST_website}
\begin{align}
\int_{-\infty}^\infty e^{-i\omega s}e^{ix\sinh as}\,ds 
= 
\frac{2}{a} e^{\pi\omega/2a}K_{i\omega/a}(x)\ \textrm{for}\ x>0\,,  
\end{align}
 and its complex conjugate, where $K_\nu(x)$ is the modified Bessel function 
 of the second kind, and recalling that $K_\nu(x)=K_{-\nu}(x)$, we find
\begin{widetext}
\begin{align}
    u^{(\omega,\mathbf{k}_\perp)}(x)
    & = \begin{cases} 2e^{\pi\omega/2a} K_{i\omega/a}
    \left(\frac{k_\perp}{a}e^{a\xi}\right)
    e^{-i\omega\tau+i\mathbf{k}_\perp\cdot\mathbf{x}_\perp}\,, 
    & \textrm{in the right Rindler wedge}, \\
     2e^{- \pi\omega/2a} K_{i\omega/a}
     \left(\frac{k_\perp}{a}e^{a\tilde{\xi}}\right)
     e^{i\omega\tilde{\tau} +i\mathbf{k}_\perp\cdot\mathbf{x}_\perp}\,, 
     & \textrm{in the left Rindler wedge}.
     \end{cases}
\end{align}
This equation is valid for all real values of $\omega$. 
Now, we define the right and left Rindler modes, 
$v^{(\mathrm{R}:\omega,\mathbf{k}_\perp)}(x)$ and 
$v^{(\mathrm{L}:\omega,\mathbf{k}_\perp)}(x)$, respectively, for $\omega >0$ by
\begin{align}
    v^{(\mathrm{R}:\omega,\mathbf{k}_\perp)}(x) & = \frac{u^{(\omega,\mathbf{k}_\perp)}(x)
    - e^{-\pi\omega/a}\overline{u^{(-\omega,-\mathbf{k}_\perp)}(x)}}
    {\sqrt{1 - e^{-2\pi\omega/a}}}\,,\label{eq:right-Rindler}\\
    v^{(\mathrm{L}:\omega,\mathbf{k}_\perp)}(x) & = \frac{u^{(-\omega,\mathbf{k}_\perp)}(x)
    - e^{-\pi\omega/a}\overline{u^{(\omega,-\mathbf{k}_\perp)}(x)}}
    {\sqrt{1 - e^{-2\pi\omega/a}}}\,. \label{eq:left-Rindler}
\end{align}
Then,
\begin{align}
      v^{(\mathrm{R}:\omega,\mathbf{k}_\perp)}(x)
      & = \begin{cases}  
      \sqrt{8\sinh(\pi\omega/a)}K_{i\omega/a}
      \left(\frac{k_\perp}{a}e^{a\xi}\right)
      e^{-i\omega\tau+i\mathbf{k}_\perp\cdot\mathbf{x}_\perp}\,, 
     & \textrm{in the right Rindler wedge}, \\
     0\,,
     & \textrm{in the left Rindler wedge},\end{cases} \\
     v^{(\mathrm{L}:\omega,\mathbf{k}_\perp)}(x)
     & = \begin{cases} 
     0\,, 
     & \textrm{in the right Rindler wedge},\\
      \sqrt{8\sinh(\pi\omega/a)}K_{i\omega/a}
      \left(\frac{k_\perp}{a}e^{a\tilde{\xi}}\right)
      e^{-i\omega\tilde{\tau}+i\mathbf{k}_\perp\cdot\mathbf{x}_\perp}\,, 
    & \textrm{in the left Rindler wedge}. 
    \end{cases}
\end{align}

The right Rindler modes are positive-frequency with respect to the Killing 
vector field $\partial/\partial\tau$, whereas the left Rindler modes are positive-frequency 
with respect to the Killing vector field $\partial/\partial\tilde{\tau}$.
The Unruh modes can be expressed in terms of the Rindler modes by inverting 
the relations~\eqref{eq:right-Rindler} and \eqref{eq:left-Rindler} for $\omega > 0$
as
\begin{align}
    u^{(\omega,\mathbf{k}_\perp)}(x) 
    & = \frac{v^{(\mathrm{R}:\omega,\mathbf{k}_\perp)}(x)
    + e^{-\pi\omega/a}\overline{v^{(\mathrm{L}:\omega,-\mathbf{k}_\perp)}(x)}}
    {\sqrt{1 - e^{-2\pi\omega/a}}}\,,\label{eq:right-Unruh}\\
    u^{(-\omega,\mathbf{k}_\perp)}(x) 
    & = \frac{v^{(\mathrm{L}:\omega,\mathbf{k}_\perp)}(x)
    + e^{-\pi\omega/a}\overline{v^{(\mathrm{R}: \omega,-\mathbf{k}_\perp)}(x)}}
    {\sqrt{1 - e^{-2\pi\omega/a}}}\,. 
    \label{eq:left-Unruh}
\end{align}
The scalar field $\widehat{\phi}(x)$ can be written as
\begin{align}
    \widehat{\phi}(x) 
    = 
    \widehat{\phi}_{\mathrm{R}}(x) + \widehat{\phi}_{\mathrm{L}}(x)\,,
\end{align}
where
\begin{equation}
    \widehat{\phi}_{\mathrm{R}}(x) 
    = 
    \int \frac{d^2\mathbf{k}_\perp}{16\pi^3 a} 
    \int_{0}^\infty d\omega  
    \left[ 
    v^{(\mathrm{R}:\omega,\mathbf{k}_\perp)}(x) 
    \hat{b}_{(\mathrm{R}:\omega,\mathbf{k}_\perp)} 
    + 
    \overline{v^{(\mathrm{R}:\omega,\mathbf{k}_\perp)}(x)}
    \hat{b}^\dagger_{(\mathrm{R}:\omega,\mathbf{k}_\perp)}
    \right]\,,
    \label{eq:phi-expansion-in-Rindler}
\end{equation}
\end{widetext}
and similarly for $\widehat{\phi}_{\mathrm{L}}(x)$. By substituting 
Eqs.~\eqref{eq:right-Unruh} and \eqref{eq:left-Unruh} into 
Eq.~\eqref{eq:phi-expansion-in-omega} and comparing the resulting 
expression with Eq.~\eqref{eq:phi-expansion-in-Rindler}, the annihilation 
operators $\hat{b}_{(\mathrm{R}:\omega,\mathbf{k}_\perp)}$ and
$\hat{b}_{(\mathrm{L}:\omega,\mathbf{k}_\perp)}$ can be expressed as
\begin{align}
  \hat{b}_{(\mathrm{R}:\omega,\mathbf{k}_\perp)} 
  & 
  = \frac{\hat{b}_{(\omega,\mathbf{k}_\perp)} 
  - e^{-\pi\omega/a}\hat{b}^\dagger_{(-\omega,-\mathbf{k}_\perp)}}
  {\sqrt{1-e^{-2\pi\omega/a}}}\,, 
  \label{eq:b-r-Rindler}\\
  \hat{b}_{(\mathrm{L}:\omega,\mathbf{k}_\perp)} 
  & 
  = \frac{\hat{b}_{(-\omega,\mathbf{k}_\perp)} 
  - e^{-\pi\omega/a}\hat{b}^\dagger_{(\omega,-\mathbf{k}_\perp)}}
  {\sqrt{1-e^{-2\pi\omega/a}}}\,. \label{eq:b-l-Rindler}
\end{align}
All right Rindler operators, $\{ \hat{b}_{(\mathrm{R}:\omega,\mathbf{k}_\perp)}, \hat{b}_{(\mathrm{R}:\omega,\mathbf{k}_\perp)}^\dagger \}$,
commute with all left Rindler operators, 
$\{ \hat{b}_{(\mathrm{L}:\omega,\mathbf{k}_\perp)}, \hat{b}_{(\mathrm{L}:\omega,\mathbf{k}_\perp)}^\dagger \}$,
and these operators satisfy 
\begin{align}
    [ \hat{b}_{(\mathrm{R}:|\omega|,\mathbf{k}_\perp)}, \hat{b}^{\dagger}_{(\mathrm{R}:|\omega'|,\mathbf{k}'_\perp)}] & = 
     [ \hat{b}_{(\mathrm{L}:|\omega|,\mathbf{k}_\perp)}, \hat{b}^{\dagger}_{(\mathrm{L}:|\omega'|,\mathbf{k}'_\perp)}] \notag \\
   &  = 8\pi^2 a\,\delta(|\omega|-|\omega'|)\delta^{(2)}(\mathbf{k}_\perp-\mathbf{k}'_\perp)\,.
   \label{eq:commutation-relations-for-LR}
\end{align}

The Fulling vacuum, $\ket{0_\mathrm{F}}$, is defined by
\begin{align}
     \hat{b}_{(\mathrm{R}:\omega,\mathbf{k}_\perp)}\ket{0_\mathrm{F}}
     =  \hat{b}_{(\mathrm{L}:\omega,\mathbf{k}_\perp)}\ket{0_\mathrm{F}} = 0\,.
\end{align}
Thus, there are no ``Rindler particles'', i.e.\ particles created by the operators $\hat{b}_{(\mathrm{R}:\omega,\mathbf{k}_\perp)}^\dagger$
or $\hat{b}_{(\mathrm{L}:\omega,\mathbf{k}_\perp)}^\dagger$, in this state.
Then, we find, from  Eq.~\eqref{eq:b-r-Rindler},
\begin{eqnarray}
\bra{0_\mathrm{M}}\hat{b}^\dagger_{(\mathrm{R}:\omega,\mathbf{k}_\perp)}
\hat{b}_{(\mathrm{R}:\omega',\mathbf{k}_\perp')}\ket{0_\mathrm{M}}
&=&
\frac{8\pi^2 a}{e^{2\pi\omega/a} -1}
\nonumber \\
&& \times
\delta(\omega-\omega')\delta^{(2)}(\mathbf{k}_\perp-\mathbf{k}_\perp')\,. \nonumber \\
\label{eq:thermal-Rindler}
\end{eqnarray}
Now, let us define the operator $\hat{A}$  by
\begin{align}
    \hat{A} = \frac{1}{\sqrt{8\pi^2 a}}\int_{0}^\infty 
    d\omega\int d^2\mathbf{k}_\perp F(\omega,\mathbf{k}_\perp)
    \hat{b}_{(\mathrm{R}:\omega,\mathbf{k}_\perp)}\,.
\end{align}
where $F(\omega,\mathbf{k}_\perp)$ is a continuous and 
square-integrable function.  Then
\begin{align}
    [ \hat{A},\hat{A}^\dagger ] 
    = 
    \int_0^\infty d\omega\int d^2\mathbf{k}_\perp\,
    |F(\omega,\mathbf{k}_\perp)|^2\,,
\end{align}
and
\begin{align}
    \bra{0_\mathrm{M}}\hat{A}^\dagger\hat{A}\ket{0_\mathrm{M}} = 
    \int_0^\infty d\omega\int d^2\mathbf{k}_\perp\, 
    \frac{|F(\omega,\mathbf{k}_\perp)|^2}
    {e^{2\pi \omega/a} - 1}\,.
\end{align}
Thus, the expected number of the ``Rindler particles'' with given Rindler energy 
(i.e.\ the energy associated with the boost Killing vector field
$\partial/\partial \tau$) is given by the Bose-Einstein distribution 
function with temperature $a/2\pi$. 
This is a manifestation of the well-known Unruh effect:
the Minkowski vacuum, $\ket{0_\mathrm{M}}$,
is a thermal state with temperature $a/2\pi$ 
with respect to the energy corresponding to the boost Killing
vector field, $\partial/\partial\tau$,  i.e.\ the Rindler energy, in the right Rindler wedge. 

     \section{Radiation by a classical source through the Unruh effect}
     \label{sec:clas_rad}

In the Heisenberg picture, analysis of radiation from a classical source 
does not depend on the quantum state because it reduces to solving the 
classical field equation (see Appendix~\ref{app-A}). Thus, the radiation can be discussed  in the 
purely classical context, as is well known. However, perhaps surprisingly, 
to analyze such a process of radiation in the interaction picture in the 
Rindler frame, it is crucial to take the Unruh effect into account as 
the present authors have emphasized in the context of the uniformly 
accelerated source.  In this section we explain this fact for 
a general classical source~\footnote{In fact, traces of the Unruh effect 
can be found in the classical field by decomposing it in terms of the 
Rindler modes~\cite{higuchi_1993}.}.

The classical source is introduced by adding the following term in the 
Lagrangian density:
\begin{align}
    \widehat{\mathcal{L}}_{\textrm{int}}(x) = j(x)\widehat{\phi}(x)\,,
\end{align}
with the corresponding interaction action,
\begin{align}
    \widehat{S}_{\textrm{int}} = \int \widehat{\mathcal{L}}_{\textrm{int}}(x)\,d^4x\,.
\end{align}
Then, in the interaction picture, this interaction adds to the final 
state the following one-particle part:
\begin{align}
    \ket{1\mathrm{p}} 
    & = i\int j(x)\widehat{\phi}(x)\ket{0_\mathrm{M}}\,d^4 x \label{eq:1-particle-state0} \\
    & = i\int \frac{d^3\mathbf{k}}{(2\pi)^3 2k} 
    \tilde{\jmath}(k)\hat{a}_\mathbf{k}^\dagger 
    \ket{0_\mathrm{M}}\,, 
    \label{eq:1-particle-state}
\end{align}
where $\tilde{\jmath}(k)$ 
is the $4$-dimensional Fourier transform of $j(x)$:
\begin{align}
    \tilde{\jmath}(k) = \int j(x)e^{ikt-i{\bf k}\cdot {\bf x}}\,d^4x\,. 
    \label{eq:Fourier-tr-of-j}
\end{align}
The final state to first order is
\begin{align}
    \ket{f} & = \left( 1 + i\mathcal{A}_{\mathrm{for}}\right)\ket{0_\mathrm{M}}
    + \ket{1\mathrm{p}}\,, \label{eq:final-state}
\end{align}
where the forward-scattering amplitude $\mathcal{A}_{\mathrm{for}}$ satisfies
\begin{align}
    2\,\mathrm{Im}\,\mathcal{A}_{\mathrm{for}} = \braket{1\mathrm{p}|1\mathrm{p}}\,,
\end{align}
so that $\braket{f|f}=1$ to first order.

The emission probability can be found as an integral over $\mathbf{k}$:
\begin{align}
    P_{\mathrm{em}}^{(\mathrm{M})} 
    & = \braket{1\mathrm{p}|1\mathrm{p}}\notag \\
    & = \int \frac{d^3\mathbf{k}}{(2\pi)^32k}|\tilde{\jmath}
    (k)|^2\,. 
    \label{eq:Mink-emission-prob}
\end{align}
The Minkowski particle number operator is defined as
\begin{align}
\widehat{N} & = \int \frac{d^3\mathbf{k}}{(2\pi)^32k}\hat{a}_\mathbf{k}^\dagger \hat{a}_\mathbf{k} \notag \\
& = 
\frac{1}{8\pi^2 a}
\int_{-\infty}^\infty d\omega \int d^2\mathbf{k}_\perp
\hat{b}^\dagger_{(\omega,\mathbf{k}_\perp)}\hat{b}_{(\omega,\mathbf{k}_\perp)}\,.
\label{eq:Minkowski-number-operator}
\end{align}
The first expression for $\widehat{N}$ can be used to show that
$\bra{f}\widehat{N}\ket{f} = \braket{1\mathrm{p}|1\mathrm{p}} = P_{\mathrm{em}}^{(\mathrm{M})} $.  That is, 
the one-particle emission probability is equal to the expected Minkowski-particle number in the final state at first order in perturbation theory.  
Equation~\eqref{eq:Minkowski-number-operator} can be regarded as a classical
result in the following sense: If we Fourier-transform the energy density
of the emitted wave, divide each $\mathbf{k}$-component by $k$ and integrate
the result over $\mathbf{k}$, then we obtain this expression~\cite{higuchi_1993}.

Now, if the classical source $j(x)$ has support only in the right 
Rindler wedge, then the same process can be described as follows.  
The field $\widehat{\phi}(x)$ restricted to the right Rindler wedge is the field
$\widehat{\phi}_{\mathrm{R}}(x)$ given by Eq.~\eqref{eq:phi-expansion-in-Rindler}.
Recall that the Minkowski vacuum is seen as a thermal bath of 
temperature $a/2\pi$ with respect to the energy associated with the 
Killing vector field $\partial/\partial\tau$.   We define the 
emission and absorption amplitudes by
\begin{align}
    \mathcal{A}^{(\omega,\mathbf{k}_\perp)}_{\mathrm{em}}
    & =  \bra{0_{\mathrm{F}}} 
    \hat{b}_{(\mathrm{R}:\omega,\mathbf{k}_\perp)}\widehat{S}_{\textrm{int}}
    \ket{0_{\mathrm{F}}} \notag \\
    & = \frac{1}{2\pi}\int j(x)\overline{v^{(\mathrm{R}:\omega,\mathbf{k}_\perp)}(x)}
    \sqrt{-g}\,d^4x\,,\\
     \mathcal{A}^{(\omega,\mathbf{k}_\perp)}_{\mathrm{abs}}
    & =  \bra{0_{\mathrm{F}}}\widehat{S}_{\textrm{int}}
    \hat{b}_{(\mathrm{R}:\omega,\mathbf{k}_\perp)}^\dagger
    \ket{0_{\mathrm{F}}} \notag \\
    & = \frac{1}{2\pi}\int j(x)v^{(\mathrm{R}:\omega,\mathbf{k}_\perp)}(x)
    \sqrt{-g}\,d^4x\,,
\end{align}
respectively, where $g$ is the determinant of the metric $g_{\mu\nu}$ in Minkowski 
(and Rindler) spacetime.  If the initial state was the Fulling vacuum, then the
emission probability would be found using the expansion~\eqref{eq:phi-expansion-in-Rindler}
of the field $\widehat{\phi}_{\mathrm{R}}$ and the commutation relation~\eqref{eq:b-commutator},
with $\hat{b}_{(\omega,\mathbf{k}_\perp)}$ replaced by $\hat{b}_{(R:\omega,\mathbf{k}_\perp)}$:
\begin{align}
    P^{(\mathrm{R},0)}_{\textrm{em}} & = \left\| \widehat{S}_{\textrm{int}}\ket{0_{\mathrm{F}}}\right\|^2 \notag \\
    & =\frac{1}{8\pi^2 a}\int_0^\infty d\omega\int d^2\mathbf{k}_\perp 
    \left|\mathcal{A}^{(\omega,\mathbf{k}_{\perp})}_{\mathrm{em}}\right|^2\,.
\end{align}
Since the Minkowski vacuum is the thermal state of temperature $a/2\pi$ with respect to the
Rindler energy,  the interaction probability for the Minkowski vacuum is
\begin{align}
    P_{\textrm{int}}^{(\mathrm{R})}
    & = \frac{1}{8\pi^2 a}\int_0^\infty d\omega\int d^2\mathbf{k}_\perp 
    \left( 
    \frac{\left|\mathcal{A}^{(\omega,\mathbf{k}_{\perp})}_{\mathrm{em}}\right|^2}
    {1-e^{-2\pi\omega/a}} 
    + \frac{\left| \mathcal{A}^{(\omega,\mathbf{k}_{\perp})}_{\mathrm{abs}}\right|^2}
    {e^{2\pi\omega/a}-1} \right)\,.
    \label{eq:P-int}
\end{align}
The first and second terms in the integrand above represent the 
(spontaneous and induced) emission 
[with $(1-e^{-2\pi\omega/a})^{-1} = 1 + (e^{2\pi\omega/a} - 1)^{-1}$] 
and absorption, respectively.

Now, as Unruh and Wald have shown~\cite{unruh-wald_1984}, both the 
emission and absorption of a particle in the thermal bath of temperature 
$a/2\pi$ in the Rindler wedge are seen as emission of a particle in Minkowski spacetime.
Hence, we expect that 
$P^{(\mathrm{R})}_{\mathrm{int}} = P^{(\mathrm{M})}_{\mathrm{em}}$.
To demonstrate this equality, we use the expansion of
$\widehat{\phi}(x)$ in terms of the Unruh modes, i.e.\ 
Eq.~\eqref{eq:phi-expansion-in-omega}, in  Eq.~\eqref{eq:1-particle-state0}.
Thus, we find
\begin{align}
    \ket{1\mathrm{p}} 
    & = \frac{i}{16\pi^3 a}
    \int_{-\infty}^\infty d\omega\int d^2\mathbf{k}_\perp
    \, \tilde{\jmath}^{\; (\mathrm{R})}(\omega,\mathbf{k}_\perp) 
    \hat{b}_{(\omega,\mathbf{k}_\perp)}^\dagger 
    \ket{0_\mathrm{M}}\,,
\end{align}
where
\begin{align}
    \tilde{\jmath}^{\; (\mathrm{R})}(\omega,\mathbf{k}_\perp)
    & = \int j(x)\overline{u^{(\omega,\mathbf{k}_\perp)}(x)}\sqrt{-g}\,d^4x\,.
\end{align}
Since the classical source $j(x)$ has support only in the right Rindler 
wedge by assumption, we have from Eqs.~\eqref{eq:right-Unruh} 
and~\eqref{eq:left-Unruh},
\begin{align}
    \tilde{\jmath}^{\; (\mathrm{R})}(\omega,\mathbf{k}_\perp)
    & = \frac{2\pi}{\sqrt{1-e^{-2\pi\omega/a}}}
    \mathcal{A}^{(\omega,\mathbf{k}_\perp)}_{\mathrm{em}}\,,\\
    \tilde{\jmath}^{\; (\mathrm{R})}(-\omega,\mathbf{k}_\perp)
    & = \frac{2\pi}{\sqrt{e^{2\pi\omega/a}-1}}
    \mathcal{A}^{(\omega,-\mathbf{k}_\perp)}_{\mathrm{abs}}\,,
\end{align}
with $\omega > 0$.
Thus, we find
\begin{eqnarray}
    \ket{1\mathrm{p}} 
    & = & 
    \frac{i}{8\pi^2 a}
    \int_0^\infty d\omega\int d^2\mathbf{k}_\perp
    \left[ \frac{\mathcal{A}^{(\omega,\mathbf{k}_\perp)}_{\mathrm{em}}}
    {\sqrt{1-e^{-2\pi\omega}}}\hat{b}_{(\omega,\mathbf{k}_\perp)}^\dagger \right.
    \nonumber \\
    &  & +
    \left. \frac{\mathcal{A}^{(\omega,-\mathbf{k}_\perp)}_{\mathrm{abs}}}
    {\sqrt{e^{2\pi\omega}-1}}\hat{b}_{(-\omega,\mathbf{k}_\perp)}^\dagger\right]
    \ket{0_{\mathrm{M}}}\,.
\end{eqnarray}
Then, we have $\braket{1\mathrm{p}|1\mathrm{p}} = P^{(\mathrm{R})}_{\mathrm{int}}$
by Eq.~\eqref{eq:b-commutator}, where $P^{(\mathrm{R})}_{\mathrm{int}}$ is 
given by Eq.~\eqref{eq:P-int}. Thus, we indeed find 
$P_{\mathrm{int}}^{(\mathrm{R})} = P^{(\mathrm{M})}_{\mathrm{em}}$ 
given by Eq.~\eqref{eq:Mink-emission-prob}.

\section{The electromagnetic and gravitational cases}
\label{sec:em_gr}

In this section we briefly discuss the electromagnetic and gravitational 
fields coupled to a classical source and show that the results for the 
massless scalar field presented in the previous sections hold for these 
fields as well. This will be the generalization of some results for the 
uniformly accelerated sources coupled to the electromagnetic 
field~\cite{vacalis_2023} and gravitational field~\cite{brito_2024} 
(see also Ref.~\cite{portales_oliva_2024}).

     \subsection{The electromagnetic case}

The Lagrangian density with the interaction term between the quantum 
field and a classical current is given by
\begin{align}
    \mathcal{L}_{\mathrm{int}} 
    = \sqrt{-g}
    \left[ - \frac{1}{4}\widehat{F}_{\mu\nu}\widehat{F}^{\mu\nu} 
    - j^\mu(x) \widehat{A}_\mu(x)
    \right]\,,
\end{align}
where $\widehat{F}_{\mu\nu} = \nabla_\mu \widehat{A}_\nu - \nabla_{\nu}\widehat{A}_\mu$.
We define the anti-symmetric tensors $\epsilon^{(\parallel)}_{\mu\nu}$ 
and $\epsilon^{(\perp)}_{\mu\nu}$ by
\begin{align}
    \epsilon^{(\parallel)}_{tz} = - \epsilon^{(\parallel)}_{zt} = 1\,,\\
    \epsilon^{(\perp)}_{xy} = - \epsilon^{(\perp)}_{yx} = 1\,,
\end{align}
with all other components vanishing.  In Rindler coordinates we have
\begin{align}
    \epsilon^{(\parallel)}_{\tau\xi} 
    = - \epsilon^{(\parallel)}_{\xi\tau} = e^{2a\xi}\,,
\end{align}
with all other components vanishing.

Two physical modes with momentum $\mathbf{k}$ can be chosen as
\begin{align}
    A_\mu^{(\mathrm{I}:\mathbf{k})}(x) 
    &  = \frac{1}{k_\perp}\epsilon^{(\parallel)}_{\mu\nu}
    \nabla^\nu f^{\mathbf{k}}(x)\,,\\
    A_\mu^{(\mathrm{I\!I}:\mathbf{k})}(x) 
    & = \frac{1}{k_\perp}\epsilon^{(\perp)}_{\mu\nu}
    \nabla^\nu f^{\mathbf{k}}(x)\,.
\end{align}
These modes are transverse, $\nabla^\mu A_\mu^{(\mathrm{P}:\mathbf{k})} = 0$, and satisfy the orthonormality condition
 for the transverse modes,
\begin{align}
     - i \int_\Sigma \overline{A^{(\mathrm{P}':\mathbf{k}')\mu}}\stackrel{\leftrightarrow}{\nabla}_\nu A^{(\mathrm{P}:\mathbf{k})}_\mu\,d\Sigma^\nu
     = (2\pi)^32k\delta^{\mathrm{P'P}}\delta^{(3)}(\mathbf{k}'-\mathbf{k})\,,  \label{eq:orthonormalEM}
\end{align}
where $\stackrel{\leftrightarrow}{\nabla}_\nu = \stackrel{\leftarrow}{\nabla}_\nu - \stackrel{\rightarrow}{\nabla}_\nu$ and 
where $\Sigma$ is a $t$=constant Cauchy surface.
Then, the quantum electromagnetic field $\widehat{A}_\mu(x)$ with 
complete gauge fixing is
\begin{align}
    \widehat{A}_\mu(x)   
    = \int \frac{d^3\mathbf{k}}{(2\pi)^32k}
    \sum_{\mathrm{P}=\mathrm{I},\mathrm{I\!I}}
    \left[ 
    A_\mu^{(\mathrm{P}:\mathbf{k})}(x)
    \,\hat{a}_{(\mathrm{P}:\mathbf{k})}
    +  
    \overline{A_\mu^{(\mathrm{P}:\mathbf{k})}(x)}\,
    \hat{a}^\dagger_{(\mathrm{P}:\mathbf{k})}
     \right]\,,
\end{align}
where the operators $\hat{a}_{(\mathrm{P}:\mathbf{k})}$ 
and $\hat{a}^\dagger_{(\mathrm{P}:\mathbf{k})}$,
$\mathrm{P}=\mathrm{I},\mathrm{I\!I}$, satisfy
\begin{align}
    \left[ 
    \hat{a}_{(\mathrm{P}:\mathbf{k})}, 
    \hat{a}^\dagger_{(\mathrm{P}':\mathbf{k}')}
    \right]
    = 
    (2\pi)^3 \, 2k \, \delta_{\mathrm{PP'}}\, 
    \delta^{(3)}(\mathbf{k}-\mathbf{k}')\,,
    \end{align}
    with all other commutators among them vanishing,  which is a consequence
    of the orthonormality condition~\eqref{eq:orthonormalEM}. The Minkowski vacuum 
    state $\ket{0_\mathrm{M}}$ is annihilated by 
    $\hat{a}_{(\mathrm{P}:\mathbf{k})}$, i.e.\ 
    $\hat{a}_{(\mathrm{P}:\mathbf{k})}\ket{0_\mathrm{M}}=0$ 
    for all $\mathrm{P}$ and $\mathbf{k}$.

The one-photon state which the classical current $j^\mu(x)$ generates is
\begin{align}
    \ket{1\mathrm{p}} 
    & = -i\int j^\mu(x)\widehat{A}_\mu(x)\ket{0_\mathrm{M}}d^4 x 
    \notag \\
    & = - \int \frac{d^3\mathbf{k}}{(2\pi)^22k}
    \tilde{\jmath}^{\,\mu} (k) k_\perp^{-1}
    \left[
    \epsilon^{(\parallel)}_{\mu\nu}k^\nu 
    \hat{a}_{(\mathrm{I}:\mathbf{k})}^\dagger 
    + \epsilon^{(\perp)}_{\mu\nu}k^\nu 
    \hat{a}_{(\mathrm{I\!I}:\mathbf{k})}^\dagger
    \right]
    \ket{0_\mathrm{M}}\,,
\end{align}
where $\tilde{\jmath}^{\,\mu}(k)$ is the Fourier transform of
$j^{\,\mu}(x)$ defined in the same way as $\tilde{\jmath}(k)$ is defined
from $j(x)$ in Eq.~\eqref{eq:Fourier-tr-of-j}.
Then, the emission probability is
\begin{align}
    P_{\mathrm{em}}^{(\mathrm{M})} & = \braket{1\mathrm{p}|1\mathrm{p}}\notag \\
    & =
    \int \frac{d^3\mathbf{k}}{(2\pi)^32k}
    \overline{\tilde{\jmath}^{\,\mu}(k)}k_\perp^{-2}
    \left[ 
    \epsilon_{\mu\alpha}^{(\parallel)}k^\alpha\epsilon_{\nu\beta}^{(\parallel)}k^\beta
    + \epsilon_{\mu\alpha}^{(\perp)}k^\alpha\epsilon_{\nu\beta}^{(\perp)}k^\beta
    \right] 
    \tilde{\jmath}^{\,\nu}(k)\,.
\end{align}
One can verify the following formula (for $k_\perp\neq 0$), e.g., in the 
Lorentz frame with $k^t=k^x=k_\perp$ and $k^z=k^y=0$:
\begin{align}
    \epsilon_{\mu\alpha}^{(\parallel)}k^\alpha\epsilon_{\nu\beta}^{(\parallel)}k^\beta
    + \epsilon_{\mu\alpha}^{(\perp)}k^\alpha\epsilon_{\nu\beta}^{(\perp)}k^\beta
    & = - k_\perp^2 g_{\mu\nu} + \frac{1}{2}(k_\mu \check{k}_\nu + k_\nu \check{k}_\mu)\,,
\end{align}
where the vector $\check{k}^\mu$ is obtained by multiplying the 
$x$- and $y$-components of $k^\mu$ by $-1$.  
Since $k_\mu \tilde{\jmath}^{\,\mu}(k) =0$, we find
\begin{align}
    \braket{1\mathrm{p}|1\mathrm{p}} & =
    -\int \frac{d^3\mathbf{k}}{(2\pi)^32k}
    \overline{\tilde{\jmath}^{\,\mu}(k)}\tilde{\jmath}_\mu(k)\,,
\end{align}
as is well known.

We define the Rindler modes by
\begin{align}
    A_\mu^{(\mathrm{RI}:\omega,\mathbf{k}_\perp)}(x) 
    &  = \frac{1}{k_\perp}\epsilon^{(\parallel)}_{\mu\lambda}
    \nabla^\lambda v^{(\mathrm{R}:\omega,\mathbf{k}_\perp)}(x)\,,\\
    A_\mu^{(\mathrm{RI\!I}:\omega,\mathbf{k}_\perp)}(x) 
    & = \frac{1}{k_\perp}\epsilon^{(\perp)}_{\mu\lambda}
    \nabla^\lambda v^{(\mathrm{R}:\omega,\mathbf{k}_\perp)}(x)\,.
\end{align}
Since the differential operators $\epsilon^{(\parallel)}_{\mu\lambda}\nabla^\lambda$ 
and  $\epsilon^{(\perp)}_{\mu\lambda}\nabla^\lambda$ commute with the operations 
for defining the Unruh and Rindler modes from the modes $f^\mathbf{k}(x)$, 
the relationship between the right Rindler modes 
$A_\mu^{(\mathrm{RP}:\omega,\mathbf{k}_\perp)}(x)$,
$\mathrm{P}=\mathrm{I}, \mathrm{I\!I}$, 
and the Minkowski modes $A_\mu^{(\mathrm{P}:\mathbf{k})}(x)$ is
exactly the same as that between $v^{(\mathrm{R}:\omega,\mathbf{k}_\perp)}(x)$ 
and $f^{\mathbf{k}}(x)$.  Hence, the quantum field $\widehat{A}_\mu(x)$ 
in the right Rindler wedge can be expanded in the same way as in the 
scalar case~\eqref{eq:phi-expansion-in-Rindler}:
\begin{eqnarray}
    \widehat{A}_{\mathrm{R}\,\mu}(x) 
    &=& 
    \int \frac{d^2\mathbf{k}_\perp}{16\pi^3a}
    \int_0^\infty d\omega 
    \sum_{\mathrm{P}=\mathrm{I},\mathrm{I\!I}}
    \left[ 
    A_\mu^{(\mathrm{RP}:\omega,\mathbf{k}_\perp)}(x)\,
    \hat{b}_{(\mathrm{RP}:\omega,\mathbf{k}_\perp)}  
    \right.
    \nonumber \\
    &+& \left. \overline{A_\mu^{(\mathrm{RP}:\omega,\mathbf{k}_\perp)}(x)}\,
    \hat{b}^\dagger_{(\mathrm{RP}:\omega,\mathbf{k}_\perp)}  
    \right]\,,
\end{eqnarray}
where the operators $\hat{b}_{(\mathrm{RP}:\omega,\mathbf{k}_\perp)}$
and $\hat{b}^\dagger_{(\mathrm{RP}:\omega,\mathbf{k}_\perp)}$ 
satisfy the commutation relations
\begin{align}
\left[\hat{b}_{(\mathrm{RP}:\omega,\mathbf{k}_\perp)},
\hat{b}^\dagger_{(\mathrm{RP'}:\omega',\mathbf{k}_\perp')}\right]
= 
8\pi^2 a\,\delta_{\mathrm{PP'}}\delta(\omega-\omega')
\delta^{(2)}(\mathbf{k}_\perp - \mathbf{k}_\perp')\,,
\end{align}
with all other commutators among them vanishing.  Then, exactly as 
in the scalar case, we find
\begin{align}
    \braket{1\mathrm{p}|1\mathrm{p}}
    & 
    = \int  \frac{d^2\mathbf{k}_\perp}{8\pi^2 a}  
    \int_0^\infty d\omega
    \sum_{\mathrm{P}=\mathrm{I},\mathrm{I\!I}}
    \left( 
    \frac{\left|\mathcal{A}^{(\mathrm{P}:\omega,\mathbf{k}_{\perp})}_{\mathrm{em}}\right|^2}
    {1-e^{-2\pi\omega/a}} 
    + \frac{\left| \mathcal{A}^{(\mathrm{P}:\omega,\mathbf{k}_{\perp})}_{\mathrm{abs}}\right|^2}
    {e^{2\pi\omega/a}-1} 
    \right) \,,
    \label{eq:vector-prob}
\end{align}
where
\begin{eqnarray}
    \mathcal{A}^{(\mathrm{P}:\omega,\mathbf{k}_\perp)}_{\mathrm{em}}
    & = &
    \frac{1}{2\pi}\int j^\mu(x)\overline{A_\mu^{(\mathrm{RP}:\omega,\mathbf{k}_\perp)}(x)}
    \sqrt{-g}\,d^4x\,,
   \\
     \mathcal{A}^{(\mathrm{P}:\omega,\mathbf{k}_\perp)}_{\mathrm{abs}}
    & = &\frac{1}{2\pi}\int j^\mu(x)A^{(\mathrm{RP}:\omega,\mathbf{k}_\perp)}_\mu(x)
    \sqrt{-g}\,d^4x\,,
\end{eqnarray}
with the same interpretation of Eq.~\eqref{eq:vector-prob} in terms of 
the Unruh effect as in the scalar case.

     \subsection{The gravitational case}

The Lagrangian density $\mathcal{L}_{\textrm{EH}}^{(2)}$ for the linearized 
gravitational field coupled to a classical stress-energy tensor is
\begin{eqnarray}
    \frac{1}{\sqrt{-g}}\mathcal{L}_{\textrm{EH}}^{(2)} 
    & = &   \frac{1}{2}\nabla_\alpha h_{\mu\nu}\nabla^\alpha h^{\mu\nu} 
    - \nabla_\alpha h_{\beta\mu}\nabla^\beta h^{\alpha\mu}\notag 
    \nonumber \\
   & + & \left(\nabla_\alpha h^{\mu\alpha} 
   - \frac{1}{2}\nabla^\mu h\right)\nabla_\mu h
    + \kappa T^{\mu\nu}h_{\mu\nu}, 
\end{eqnarray}
where $h = h^\alpha{}_\alpha$ and $\kappa = \sqrt{8\pi G}$.  Two physical modes 
with momentum $\mathbf{k}$ can be chosen as
\begin{equation}
 h^{(\mathrm{I}:\mathbf{k})}_{\mu\nu}(x) 
 = 
 \frac{1}{\sqrt{2}}\left[ g_{\mu\nu} 
 + 2q_{\mu\nu}(\mathbf{k}_\bot)\right]f^{\mathbf{k}}(x)\,,
\end{equation}
where
\begin{equation}
q_{\mu\nu}(\mathbf{k}_\bot)  
= 
\begin{cases}
-g_{\mu\nu} - {k_\mu k_\nu}/{k_\perp^2}\,, & \textrm{if}\ \mu,\nu=x\ \textrm{or}\ y,\\
 0\,, & \textrm{otherwise}\,,
\end{cases}
\end{equation}
and
\begin{align}
     h^{(\mathrm{I\!I}:\mathbf{k})}_{\mu\nu}(x)
     = \frac{1}{\sqrt{2}k_\perp^2}
     \left[ 
     \epsilon^{(\parallel)}_{\mu\alpha}\epsilon^{(\perp)}_{\nu\beta}
      + \epsilon^{(\parallel)}_{\nu\alpha}\epsilon^{(\perp)}_{\mu\beta} 
      \right]
      \nabla^\alpha\nabla^\beta f^{\mathbf{k}}(x)\,.
\end{align}
These modes satisfy the de~Donder condition,
\begin{align}
    \nabla^\mu h_{\mu\nu}^{(\mathrm{P}:\mathbf{k})} - \frac{1}{2}\nabla_\nu h^{(\mathrm{P}:\mathbf{k})} = 0\,, \label{eq:de-Donder}
\end{align}
and the normalization condition for the modes satisfying Eq.~\eqref{eq:de-Donder},
\begin{align} 
& i \int \left[ \overline{h^{(\mathrm{P}':\mathbf{k}')\mu\nu}}\stackrel{\leftrightarrow}{\nabla}_\alpha h^{(\mathrm{P}:\mathbf{k})}_{\mu\nu}
- \frac{1}{2}\overline{h^{(\mathrm{P}':\mathbf{k}')}}\stackrel{\leftrightarrow}{\nabla}_\alpha h^{(\mathrm{P}:\mathbf{k})}\right]d\Sigma^\alpha \notag \\
& = (2\pi)^3 2k\,\delta^{\mathrm{P'P}}\delta^{(3)}(\mathbf{k}'-\mathbf{k})\,.
\end{align}

Then, the emission probability from the classical stress-energy tensor 
$T^{\mu\nu}(x)$ is
\begin{align}
    \braket{1\mathrm{p}|1\mathrm{p}}
    & = \int \frac{d^3\mathbf{k}}{(2\pi)^32k} 
    \sum_{\mathrm{P}=\mathrm{I},\mathrm{I\!I}}
    \left|\mathcal{A}^{(\mathrm{P}:\mathbf{k})}\right|^2\,,
\end{align}
where
\begin{align}
    \mathcal{A}^{(\mathrm{P}:\mathbf{k})} = 
    \kappa\int T^{\mu\nu}(x) \overline{h^{(\mathrm{P}:\mathbf{k})}_{\mu\nu}(x)}
    \sqrt{-g}\,d^4x\,.
\end{align}
Define the Fourier transform of $T^{\mu\nu}(x)$ by
\begin{align}
    \mathcal{T}^{\mu\nu}(k) = \int T^{\mu\nu}(x)e^{ik\cdot x} d^4 x\,.
\end{align}
(Note that $\sqrt{-g}=1$ here.)
Then,
\begin{align}
\braket{1\mathrm{p}|1\mathrm{p}}
 = \kappa^2\int \frac{d^3\mathbf{k}}{(2\pi)^32k}
\overline{\mathcal{T}^{\mu\nu}(k)}
S_{\mu\nu\lambda\sigma}(k)
\mathcal{T}^{\lambda\sigma}(k)\,,
\end{align}
where
\begin{eqnarray}
S_{\mu\nu\lambda\sigma} 
& = &
\frac{1}{2}(g_{\mu\nu} + 2q_{\mu\nu})(g_{\lambda\sigma}+2q_{\lambda\sigma})
\nonumber \\
& + & \frac{2}{k_\perp^4}
\epsilon^{(\parallel)}_{\mu\alpha}k^\alpha
\epsilon^{(\perp)}_{\nu\beta}k^\beta 
\epsilon^{(\parallel)}_{\lambda\gamma} k^\gamma
\epsilon^{(\perp)}_{\sigma\delta}k^\delta\,.
\end{eqnarray}
 By working in the Lorentz frame where $k^z=k^y=0$ so that 
$k^t=k^x=k_\perp$ (with the assumption that $k_\perp\neq 0$) and using 
$k_\mu\mathcal{T}^{\mu\nu}(k)=0$, we find
\begin{align}
    \overline{\mathcal{T}^{\mu\nu}}(k)
    S_{\mu\nu\lambda\sigma}(k)
    \mathcal{T}^{\lambda\sigma}(k)
    & = 
    \overline{\mathcal{T}^{\mu\nu}(k)}\mathcal{T}_{\mu\nu}(k) 
    - \frac{1}{2}\overline{\mathcal{T}(k)}\mathcal{T}(k)\,,
\end{align}
where $\mathcal{T}(k) = \mathcal{T}^\mu{}_\mu(k)$.
Thus, the emission probability is
\begin{align}
\braket{1\mathrm{p}|1\mathrm{p}}
 = \kappa^2\int \frac{d^3\mathbf{k}}{(2\pi)^32k}
 \left[\mathcal{T}^{\mu\nu}(k)\mathcal{T}_{\mu\nu}(k) 
 - \frac{1}{2}\overline{\mathcal{T}}(k)\mathcal{T}(k)\right]\,,
\end{align}
as is well known.

The physical right Rindler modes can be chosen as
\begin{align}
 h^{(\mathrm{IR}:\omega,\mathbf{k}_\perp)}_{\mu\nu}(x) 
 & = \frac{1}{\sqrt{2}}\left[ g_{\mu\nu} 
 + 2q_{\mu\nu}(\mathbf{k}_\bot)\right] v^{(\mathrm{R}:\omega,\mathbf{k}_\perp)}(x)\,,
 \label{eq:grav-mode-1}\\
 h^{(\mathrm{I\!IR}:\omega,\mathbf{k}_\perp)}_{\mu\nu}(x)
 & = 
 \frac{1}{\sqrt{2}k_\perp^2}
 \left[ 
 \epsilon^{(\parallel)}_{\mu\alpha}\epsilon^{(\perp)}_{\nu\beta}
 + \epsilon^{(\parallel)}_{\nu\alpha}\epsilon^{(\perp)}_{\mu\beta} 
 \right]
 \nabla^\alpha\nabla^\beta  v^{(\mathrm{R}:\omega,\mathbf{k}_\perp)}(x)\,. 
 \label{eq:grav-mode-2}
\end{align}
The modes $h^{(\mathrm{I\!IR}:\omega,\mathbf{k}_\perp)}_{\mu\nu}(x)$ 
were found in Ref.~\cite{sugiyama_2021} up to a normalization factor
whereas the modes $h_{\mu\nu}^{(\mathrm{IR}:\omega,\mathbf{k}_\perp)}(x)$ 
were obtained in Ref.~\cite{brito_2024} by a gauge transformation from
the modes given in Ref.~\cite{sugiyama_2021} up to a normalization factor.
The tensors in Eq.~\eqref{eq:grav-mode-1} and the differential operator 
in Eq.~\eqref{eq:grav-mode-2} commute with the operations for defining
the Unruh and Rindler modes from the Minkowski modes.  Hence, the relationship 
between the description of the emission process in Minkowski spacetime and
the interaction of the classical source with the thermal bath is exactly 
the same as in the scalar case. Thus, just like in the scalar case, if we 
define the emission and absorption amplitudes in the right Rindler wedge as
\begin{align}
    \mathcal{A}^{(\mathrm{P}:\omega,\mathbf{k}_\perp)}_{\mathrm{em}}
    & = 
    \frac{\kappa}{2\pi}\int T^{\mu\nu}(x)
    \overline{h_{\mu\nu}^{(\mathrm{RP}:\omega,\mathbf{k}_\perp)}(x)}
    \sqrt{-g}\,d^4x\,,\\
     \mathcal{A}^{(\mathrm{P}:\omega,\mathbf{k}_\perp)}_{\mathrm{abs}}
    & = 
    \frac{\kappa}{2\pi}\int T^{\mu\nu}(x)
    h^{(\mathrm{RP}:\omega,\mathbf{k}_\perp)}_{\mu\nu}(x)
    \sqrt{-g}\,d^4x\,,
\end{align}
respectively, 
then Eq.~\eqref{eq:vector-prob} for $\braket{1\mathrm{p}|1\mathrm{p}}$  
holds as in the electromagnetic case together with its interpretation in terms of
the Unruh effect.

\section{Discussion}
\label{sec:discuss}

We have shown that what uniformly accelerated observers interpret as
interaction of an {\it arbitrary} charge distribution with
the Unruh thermal bath is interpreted by inertial observers as emission of radiation by the same charge distribution. 
All calculations were performed in first-order perturbation theory, 
which means that the inertial-frame result corresponds to {\it classical} 
radiation---once one accepts that the latter consists of {\it quanta} of 
energy satisfying Planck's energy-frequency relation. 
This result, obtained for the scalar,  electromagnetic,
and  graviton fields,  corroborates the claim that classical
radiation observed in the inertial frame can already be considered
as an ``observation'' of the Unruh effect
in the same sense that, in Newtonian mechanics, 
the planetary motion in the inertial
frame can be considered to be providing evidence of the inertial forces, e.g., the centrifugal force,
in the rotating frame.
This observation has already been made
in the literature for the case of point-like sources
in specific trajectories~\cite{higuchi_1992R, higuchi_1992, ren_1994, 
landulfo_2019, portales_oliva_2022, vacalis_2023, portales_oliva_2024, 
brito_2024}. Here, the observation is extended to arbitrary currents 
for the scalar, electromagnetic, and gravitational fields. 
In our view, the analysis presented here proves, beyond any 
 doubt, that classical radiation seen by inertial observers,
with the only extra assumption that it is constituted by {\it quanta}
of energy, is a testimony
to the existence of the Unruh thermal bath in the uniformly 
accelerated 
frame. Those who dispute this observation
would have to reproduce the radiation from the classical source
in a uniformly accelerated frame without using the Unruh effect.


\acknowledgments

    We are thankful to F.\ Portales-Oliva for calling our attention to Ref.~\cite{connel_2020}. D.~V.~would like to thank the  Institute for Quantum Optics and Quantum Information of the Austrian Academy of Science for hosting him during a sabbatical year. G.~M.~and D.~V.~were supported in part by S\~{a}o Paulo Research Foundation (FAPESP) under grants 2021/09293-7 and 2023/04827-9, respectively. G.~M.~was also partially supported by Conselho Nacional de Desenvolvimento Cientifico e Tecnologico (CNPq) under grant 302674/2018-7. The work of J.~B. and L.~C. 
    was supported in part by Funda\c{c}\~ao Amaz\^onia de Amparo a Estudos e
Pesquisas (FAPESPA), CNPq and Coordena\c{c}\~ao de Aperfei\c{c}oamento de
Pessoal de N\'ivel Superior (CAPES) -- Finance Code 001. Their work has also been supported by
the European Horizon Europe staff exchange (SE) programme
HORIZONMSCA-2021-SE-01 Grant No. NewFunFiCO-101086251.  
The work of R.~B. and G.~G. was supported in part by EPSRC grants no. EP/X01133X/1 and EP/X010791/1. G.~G. is also a member of the Quantum Sensing for the Hidden Sector (QSHS) collaboration, supported by STFC grant no. ST/T006277/1.

\appendix

\section{Radiation in the Heisenberg picture}
\label{app-A}

In this Appendix we 
discuss radiation of massless scalar field 
in $4$ dimensions from a 
classical source in the Heisenberg picture.  This confirms that the radiation
formula from a classical source does not depend on the quantum state.

The field equation for the Heisenberg operator for a massless scalar field, $\widehat{\phi}_{\mathrm{H}}(x)$, with a classical source term $j(x)$ is
\begin{align}
    \Box \widehat{\phi}_{\mathrm{H}}(x) = j(x)\,,
\end{align}
where we assume $j(x)$ to be compactly supported.  
The retarded Green's function $G_R(x,x')$ satisfying
\begin{align}
    \Box_x G_R(x,x') = \delta^{(4)}(x-x')\,,
\end{align}
is given by
\begin{align}
    G_R(x,x') = i\theta(t-t')\int \frac{d^3\mathbf{k}}{(2\pi)^3 2k}
    \left[ e^{-ik\cdot (x-x')} - e^{ik\cdot (x-x')}\right]\,,
\end{align}
where $\theta$ is the Heaviside step function.

Let $\widehat{\phi}(x)$ be the quantum field without the source $j(x)$ 
that agrees with
$\widehat{\phi}_{\mathrm{H}}(x)$ in the past of the source.  Then
the field $\widehat{\phi}_{\mathrm{H}}(x)$ at a time in the future of the source is
\begin{align}
    \widehat{\phi}_{\mathrm{H}}(x) & = \widehat{\phi}(x) + \phi^{(\mathrm{cl})}(x)\,,
\end{align}
where
\begin{align}
\phi^{(\mathrm{cl})}(x) &  = \int G_R(x,x')j(x')\,d^4x' \notag \\
    & = i \int \frac{d^3\mathbf{k}}{(2\pi)^3 2k}
    \left[ \tilde{\jmath}(k) e^{-ik\cdot x} - \overline{\tilde{\jmath}(k)}e^{ik\cdot x}\right]\,,
\end{align}
where $\tilde{\jmath}(k)$ is the Fourier transform of $j(x)$ defined by Eq.~\eqref{eq:Fourier-tr-of-j}.
Here we have used the fact that the source $j(x)$ is real.
Thus, if we expand the fields $\widehat{\phi}_{\mathrm{H}}(x)$ and $\widehat{\phi}(x)$ as
\begin{align}
    \widehat{\phi}_{\mathrm{H}}(x) & = \int\frac{d^3\mathbf{k}}{(2\pi)^32k}
    \left[ \hat{a}_{\mathbf{k}}^{\mathrm{H}} e^{-ik\cdot x} + \hat{a}_{\mathbf{k}}^{\mathrm{H}}
    e^{ik\cdot x}\right]\,,\\
    \widehat{\phi}(x) & = \int\frac{d^3\mathbf{k}}{(2\pi)^32k}
    \left[ \hat{a}_{\mathbf{k}} e^{-ik\cdot x} + \hat{a}_{\mathbf{k}}
    e^{ik\cdot x}\right]\,,
\end{align}
in the future of the source, then we have
\begin{align}
    \hat{a}^{\mathrm{H}}_{\mathbf{k}} = \hat{a}_{\mathbf{k}} + i \tilde{\jmath}(k)\,.
\end{align}

We assume that the Heisenberg state $\ket{\mathrm{H}}$ satisfies
$\bra{\mathrm{H}}a_{\mathbf{k}}\ket{\mathrm{H}} = 0$ for all $\mathbf{k}$.
The initial and final number operators, $\widehat{N}^{(i)}$ and $\widehat{N}^{(f)}$, are defined by
\begin{align}
    \widehat{N}^{(i)} & = \int \frac{d^3\mathbf{k}}{(2\pi)^32k}\hat{a}_{\mathbf{k}}^\dagger a_{\mathbf{k}}\,,\\
    \widehat{N}^{(f)} & = \int \frac{d^3\mathbf{k}}{(2\pi)^32k}\hat{a}_{\mathbf{k}}^{\mathrm{H}\dagger} a_{\mathbf{k}}^{\mathrm{H}}\,.
\end{align}
Then, the initial and final particle numbers, 
$\bra{\mathrm{H}}\widehat{N}^{(i)}\ket{\mathrm{H}}$ and $\bra{\mathrm{H}}\widehat{N}^{(f)}\ket{\mathrm{H}}$, are ill-defined in general, but their difference, 
i.e.\ the increase in the particle number, is well-defined and state independent.  It is given by 
\begin{align}
    \Delta N & = \bra{\mathrm{H}}\widehat{N}^{(f)}\ket{\mathrm{H}}
    - \bra{\mathrm{H}}\widehat{N}^{(i)}\ket{\mathrm{H}} \notag \\
    & = \int \frac{d^3\mathrm{k}}{(2\pi)^32k}|\tilde{\jmath}(k)|^2\,.
\end{align}
Thus, the increase in the particle number is independent of the state and agrees with
the classical result.

\section{Emission from a classical source in the Fulling vacuum}
\label{app-B}

In this Appendix we analyze the emission process from a classical source in the Fulling vacuum state in the interaction picture. We find that the classical 
result is reproduced differently compared to the case with the Minkowski vacuum
state.

The final state to first order in perturbation theory is
\begin{align}
    \ket{f} & = \left( 1 +i\mathcal{A}_{\mathrm{for}}\right)\ket{0_\mathrm{F}}
    + \ket{1\mathrm{p}^{(\mathrm{R})}}\,, \label{eq:final-state-Rindler}
\end{align}
where the final $1$-particle state is
\begin{align}
    \ket{1\mathrm{p}^{(\mathrm{R})}} = \frac{i}{8\pi^2 a}\int_{0}^\infty d\omega
    \int d^2\mathbf{k_\perp} \mathcal{A}^{(\omega,\mathbf{k}_\perp)}_{\mathrm{em}}\hat{b}^\dagger_{(\mathrm{R}:\omega,\mathbf{k}_\perp)}\ket{0_\mathrm{F}}\,, \label{eq:1particle-Rindler-state}
\end{align}
and where the forward-scattering amplitude $\mathcal{A}_{\mathrm{for}}$ satisfies
\begin{align}
    2\,\mathrm{Im}\,\mathcal{A}_{\mathrm{for}} = \braket{1\mathrm{p}^{(\mathrm{R})}|1\mathrm{p}^{(\mathrm{R})}}\,.
\end{align}
[See Eq.~\eqref{eq:final-state} for the case with the Minkowski vacuum.] 
We find
\begin{align}
    \braket{1\mathrm{p}^{(\mathrm{R})}|1\mathrm{p}^{(\mathrm{R})}}
    & = \frac{1}{8\pi^2 a}\int_{0}^\infty d\omega
    \int d^2\mathbf{k_\perp} \left|\mathcal{A}^{(\omega,\mathbf{k}_\perp)}_{\mathrm{em}}\right|^2\,.
\end{align}
The right Rindler particle number operator is
\begin{align}
    \widehat{N}^{(\mathrm{R})} = \int_0^\infty d\omega \int d^2\mathbf{k}_\perp
    \hat{b}^\dagger_{(\mathrm{R}:\omega,\mathbf{k}_\perp)}
     \hat{b}_{(\mathrm{R}:\omega,\mathbf{k}_\perp)}\,,
\end{align}
and similarly for the left Rindler particle number $\widehat{N}^{(\mathrm{L})}$.
Then, we have $\langle 1\mathrm{p}^{(\mathrm{R})}|\widehat{N}^{(R)}| 1\mathrm{p}^{(\mathrm{R})}\rangle = \langle 1\mathrm{p}^{(\mathrm{R})}|1\mathrm{p}^{(\mathrm{R})}\rangle$ and
$\langle 1\mathrm{p}^{(\mathrm{R})}|\widehat{N}^{(L)}| 1\mathrm{p}^{(\mathrm{R})}\rangle = 0$.
Then, by Eq.~\eqref{eq:final-state-Rindler} we find
\begin{align}
    \braket{f|\widehat{N}^{(\mathrm{R)})}|f} & = \braket{1\mathrm{p}^{(\mathrm{R})}|1\mathrm{p}^{(\mathrm{R})}},\\
    \braket{f|\widehat{N}^{(\mathrm{L)})}|f} & = 0\,.
\end{align}
Thus, the particle-emission probability is equal to the expected Rindler particle number in the final state
rather than the expected Minkowski particle number.

Now let us find the increase in the expected Minkowski particle number in the interaction picture and verify that it
agrees with the Heisenberg picture, which gives the classical result as seen in Appendix~\ref{app-A}.
The Fulling vacuum is not the Minkowski vacuum, and hence, the expected Minkowski particle number is nonzero.
(In fact it is infinite.)
First from Eqs.~\eqref{eq:b-r-Rindler} and \eqref{eq:b-l-Rindler} we find
that the Fulling vacuum $\ket{0_\mathrm{F}}$ has the following expectation values for the Unruh modes:
\begin{align}
    \bra{0_\mathrm{F}} b^\dagger_{(\omega,\mathbf{k}_\perp)}
    b_{(\omega',\mathbf{k}_\perp')}\ket{0_\mathrm{F}}
    & = \frac{8\pi^2 a}{e^{2\pi|\omega|/a} - 1}\delta(\omega-\omega')
    \delta^{(2)}(\mathbf{k}_\perp -\mathbf{k}'_\perp)\,.
\end{align}
Notice the similarity of this equation with Eq.~\eqref{eq:thermal-Rindler}.
We also find, using Eqs.~\eqref{eq:b-r-Rindler}, 
\eqref{eq:b-l-Rindler} and \eqref{eq:commutation-relations-for-LR},
\begin{align}
&  \bra{1\mathrm{p}^{(\mathrm{R})}} b^\dagger_{(|\omega|,\mathbf{k}_\perp)}
    b_{(|\omega'|,\mathbf{k}_\perp')}\ket{1\mathrm{p}^{(\mathrm{R})}} \notag \\
    & =
    \frac{\bra{1\mathrm{p}^{(\mathrm{R})}}b^\dagger_{(\mathrm{R}:|\omega|,\mathbf{k}_\perp)}
    b_{(\mathrm{R}:|\omega'|,\mathbf{k}_\perp')}\ket{1\mathrm{p}^{(\mathrm{R})}}}{\sqrt{(1-e^{-2\pi|\omega|/a})(1-e^{-2\pi|\omega'|/a})}}\notag \\
    & \qquad +
    \frac{8\pi^2 a\,\delta(|\omega|-|\omega'|)\delta^{(2)}(\mathbf{k}_\perp - \mathbf{k}_\perp')}{\sqrt{(e^{2\pi|\omega|/a}-1)(e^{2\pi|\omega'|/a}-1)}}\braket{1\mathrm{p}^{(\mathrm{R})}|1\mathrm{p}^{(\mathrm{R})}}\,,
    \label{eq:1pR-bb-1pR1}\\
  &  \bra{1\mathrm{p}^{(\mathrm{R})}} b^\dagger_{(-|\omega|,\mathbf{k}_\perp)}
    b_{(-|\omega'|,\mathbf{k}_\perp')}\ket{1\mathrm{p}^{(\mathrm{R})}} \notag \\
 & =  \frac{\bra{1\mathrm{p}^{(\mathrm{R})}}b^\dagger_{(\mathrm{R}:|\omega'|,\mathbf{k}'_\perp)}
    b_{(\mathrm{R}:|\omega|,\mathbf{k}_\perp)}\ket{1\mathrm{p}^{(\mathrm{R})}}}{\sqrt{(e^{2\pi|\omega|/a}-1)(e^{2\pi|\omega'|/a}-1)}}\notag \\
    & \qquad +
    \frac{8\pi^2 a\,\delta(|\omega|-|\omega'|)\delta^{(2)}(\mathbf{k}_\perp - \mathbf{k}_\perp')}{\sqrt{(e^{2\pi|\omega|/a}-1)(e^{2\pi|\omega'|/a}-1)}}\braket{1\mathrm{p}^{(\mathrm{R})}|1\mathrm{p}^{(\mathrm{R})}}\,.
    \label{eq:1pR-bb-1pR2}
\end{align}
Using the expression \eqref{eq:1particle-Rindler-state} for the state $\ket{1\mathrm{p}^{(\mathrm{R})}}$  and the commutation relations~\eqref{eq:commutation-relations-for-LR} we obtain
\begin{align}
    \bra{1\mathrm{p}^{(\mathrm{R})}} b^\dagger_{(\mathrm{R}:|\omega|,\mathbf{k}_\perp)}
    b_{(\mathrm{R}:|\omega'|,\mathbf{k}_\perp')}\ket{1\mathrm{p}^{(\mathrm{R})}}
     = \overline{\mathcal{A}^{(|\omega|,\mathbf{k}_\perp)}_{\mathrm{em}}}
    \mathcal{A}^{(|\omega'|,\mathbf{k}_\perp')}_{\mathrm{em}}\,.
\end{align}
Substituting this equation into Eqs.~\eqref{eq:1pR-bb-1pR1} and \eqref{eq:1pR-bb-1pR2} and using the
expression \eqref{eq:final-state-Rindler}, we find to lowest nontrivial order,
\begin{align}
    \bra{f}b^\dagger_{(|\omega|,\mathbf{k}_\perp)}
    b_{(|\omega'|,\mathbf{k}_\perp')}\ket{f} 
    - \bra{0_{\mathrm{F}}}b^\dagger_{(|\omega|,\mathbf{k}_\perp)}
    b_{(|\omega'|,\mathbf{k}_\perp')}\ket{0_{\mathrm{F}}}\notag \\
     = \frac{\overline{\mathcal{A}^{(|\omega|,\mathbf{k}_\perp)}_{\mathrm{em}}}
    \mathcal{A}^{(|\omega'|,\mathbf{k}_\perp')}_{\mathrm{em}}}
    {\sqrt{(1-e^{-2\pi|\omega|/a})(1-e^{-2\pi|\omega'|/a})}}\,,\\
    \bra{f}b^\dagger_{(-|\omega|,\mathbf{k}_\perp)}
    b_{(-|\omega'|,\mathbf{k}_\perp')}\ket{f} 
    - \bra{0_{\mathrm{F}}}b^\dagger_{(-|\omega|,\mathbf{k}_\perp)}
    b_{(-|\omega'|,\mathbf{k}_\perp')}\ket{0_{\mathrm{F}}}\notag \\
     = \frac{\mathcal{A}^{(|\omega|,\mathbf{k}_\perp)}_{\mathrm{em}}
    \overline{\mathcal{A}^{(|\omega'|,\mathbf{k}_\perp')}_{\mathrm{em}}}}
    {\sqrt{(e^{2\pi|\omega|/a}-1)(e^{2\pi|\omega'|/a}-1)}}\,.
\end{align}
Then, from the second expression in Eq.~\eqref{eq:Minkowski-number-operator} for the Minkowski particle number
operator, $\widehat{N}$, we find, using $\left|\mathcal{A}^{(|\omega|,\mathbf{k}_\perp)}_{\mathrm{em}}\right|=\left|\mathcal{A}^{(|\omega|,\mathbf{k}_\perp)}_{\mathrm{abs}}\right|$,
\begin{align}
    \bra{f}\widehat{N}\ket{f} - \bra{0_\mathrm{F}}\widehat{N}\ket{0_\mathrm{F}}
    = P_{\mathrm{int}}^{(\mathrm{R})}\,,
\end{align}
where $P_{\mathrm{int}}^{(\mathrm{R})}$ is given by Eq.~\eqref{eq:P-int},
which was shown to be equal to $P_{\mathrm{em}}^{(\mathrm{M})}$, which in turn is equal to the classical result [see Eq.~\eqref{eq:Mink-emission-prob}]. Thus, inertial observers in the Fulling vacuum will witness usual Larmor radiation being emitted from classical sources over the nontrivial particle content they experience. The corresponding Minkowski particles, as described by inertial observers, should be associated with the emission of Rindler particles, as described by uniformly accelerated observers at 0\,K.



\begin{thebibliography}{99}

\bibitem{unruh_1976}
{W.~G.~Unruh, 
Notes on black-hole evaporation, 
Phys. Rev. D \textbf{14}, 870 (1976).}

\bibitem{hawking_1974} 
{S.~W.~Hawking, 
Black hole explosions?, Nature (London) 248, 30 (1974).}

\bibitem{hawking_1976}
{Particle creation by black holes, Commun. Math. Phys. 43, 199 (1975),
[Erratum: Commun. Math. Phys. 46, 206 (1976)].}

\bibitem{bisognano_1975}
{J.~J.~Bisognano and E.~H.~Wichmann, 
On the duality condition for a Hermitian scalar field, J. Math. Phys. \textbf{16}, 985 (1975).}

\bibitem{bisognano_1976}
{J.~J.~Bisognano and E.~H.~Wichmann,
On the duality condition for quantum fields, J. Math. Phys. \textbf{17}, 303 (1976).}

\bibitem{sewell_1982} 
{G.~L.~Sewell, 
Quantum fields on manifolds: PCT and gravitationally induced thermal states, 
Ann. Phys. (N.Y.) \textbf{141}, 201 (1982).}

\bibitem{fulling_2004}
{S.~A.~Fulling and W.~G.~Unruh, 
Comment on ``Boundary conditions in the Unruh problem'', 
Phys. Rev. D \textbf{70}, 048701 (2004).}

\bibitem{belinskii_1997}
{V.~A.~Belinskii,  B.~M.~Karnakov,  V.~D.~Mur,  and N.~B.~Narozhnyi,  Does the Unruh effect exist?, Pis’ma Zh. Eksp. Teor. Fiz. \textbf{65}, 861 [JETP Lett. \textbf{65}, 902 (1997)].}

\bibitem{narozhny_2001}
N.~B.~Narozhny, A.~M.~Fedotov, B.~M.~Karnakov, V.~D.~Mur, V.~A.~Belinskii, 
Boundary conditions in the Unruh problem,  
Phys. Rev. D {\bf 65}, 025004 (2001).

\bibitem{ford_2006}
G.~W.~Ford and R.~F.~O’Connell, 
Is there Unruh radiation? 
Phys. Lett. A {\bf 350}, 17 (2006).

\bibitem{gelfer_2015}
E.~G.~Gelfer, A.~M.~Fedotov, V.~D.~Mur, and N.~B.~Narozhny,  
Boost modes for a massive fermion field and the Unruh problem,  
Theor. Math. Phys. {\bf 182}, 356 (2015).

\bibitem{cruz_2016}
S.~Cruz~y~Cruz and B.~Mielnik, Non-Inertial
Quantization: Truth or Illusion?, J. Phys. Conf. Ser. \textbf{698},
012002 (2016).

\bibitem{connel_2020}
R.~F.~O’Connell, 
Demise of Unruh radiation, 
Mod. Phys. Lett. A {\bf 35}, 2050329 (2020).

\bibitem{popruzhenko_2023}
S.~V.~Popruzhenko and A.~M.~Fedotov,
Dynamics and radiation of charged particles in ultra-intense laser fields,
Phys-Uspekhi {\bf 66}, 460 (2023).

\bibitem{bell_1983}
{J.~S.~Bell and J.~M.~Leinaas, 
Electrons as accelerated thermometers, 
Nucl. Phys. B \textbf{212}, 131 (1983).}

\bibitem{vanzella_2001}
{D.~A.~T.~Vanzella and G.~E.~A.~Matsas, 
Decay of accelerated protons and the existence of the Fulling-Davies-Unruh effect, 
Phys. Rev. Lett. \textbf{87}, 151301 (2001).}

\bibitem{suzuki_2003}
{H.~Suzuki and K.~Yamada, 
Analytic evaluation of the decay rate for an accelerated proton, 
Phys. Rev. D \textbf{67}, 065002 (2003).}

\bibitem{cozzella_2018}
{G.~Cozzella,  S.~A.~Fulling, A.~G.~S.~Landulfo, G.~E.~A.~Matsas, and D.~A.~T.~Vanzella, 
Unruh effect for mixing neutrinos, 
Phys. Rev. D \textbf{97}, 105022 (2018).}

\bibitem{chen_1999}
{P.~Chen and T.~Tajima, 
Testing Unruh radiation with ultraintense lasers, 
Phys. Rev. Lett. \textbf{83}, 256 (1999).}

\bibitem{schutzhold_2006}
{R.~Sch\"utzhold, G.~Schaller, and D.~Habs, 
Signatures of the Unruh effect from electrons accelerated by ultrastrong laser fields, 
Phys. Rev. Lett. \textbf{97}, 121302 (2006).}

\bibitem{cozzella_2017}
{G.~Cozzella, A.~G.~S.~Landulfo, G.~E.~A.~Matsas, and D.~A.~T.~Vanzella,  
Proposal for observing the Unruh effect using classical electrodynamics, 
Phys. Rev. Lett. \textbf{118}, 161102 (2017).}

\bibitem{lima_2019}
{C.~A.~U.~Lima, F.~Brito, J.~A.~Hoyos, and D.~A.~T.~Vanzella, 
Probing the Unruh effect with an accelerated extended system, 
Nature Communications \textbf{10}, 3030 (2019).}

\bibitem{lynch_2021}
{M.~H.~Lynch, E.~Cohen, Y.~Hadad, and I.~Kaminer,  
Experimental observation of acceleration-induced thermality, 
Phys. Rev. D \textbf{104}, 025015 (2021).}

\bibitem{leonhardt_2018}
{U.~Leonhardt, I.~Griniasty, S.~Wildeman, E.~Fort, and M.~Fink,  
Classical analog of the Unruh effect, 
Phys. Rev. A \textbf{98}, 022118 (2018).}

\bibitem{barros_2020}
{G.~B.~Barros, J.~P.~C.~R.~Rodrigues, A.~G.~S.~Landulfo, and G.~E.~A.~Matsas,  
Traces of the Unruh effect in surface waves, 
Phys. Rev. D \textbf{101}, 065015 (2020).}

\bibitem{larmor_1897}
{J.~Larmor, 
On the theory of the magnetic influence on spectra; and on the radiation from moving ions, 
Philos. Mag. \textbf{44}, 503 (1897).}

\bibitem{jackson_1998}
{J.~D.~Jackson, 
\textit{Classical Electrodynamics}, \textit{3rd ed. }
(Wiley, New York, 1998).}

\bibitem{higuchi_1992R}
{A.~Higuchi, G.~E.~A.~Matsas, and D.~Sudarsky, 
Bremssstrahlung and zero-energy Rindler photons, 
Phys. Rev. D \textbf{45}, R3308(R) (1992).}

\bibitem{higuchi_1992}
{A.~Higuchi, G.~E.~A.~Matsas, and D.~Sudarsky, 
Bremsstrahlung and Fulling-Davies-Unruh thermal bath, 
Phys. Rev. D \textbf{46}, 3450 (1992).}

\bibitem{ren_1994}
{H.~Ren and E.~J.~Weinberg, 
Radiation from a moving scalar source, 
Phys. Rev. D \textbf{49}, 6526 (1994).}

\bibitem{landulfo_2019}
{A.~G.~S.~Landulfo, S.~A.~Fulling,  and G.~E.~A.~Matsas, 
Classical and quantum aspects of the radiation emitted by a uniformly 
accelerated charge: Larmor-Unruh reconciliation and zero-frequency 
Rindler modes, 
Phys. Rev. D \textbf{100}, 045020 (2019).}

\bibitem{portales_oliva_2022}
{F.~Portales-Oliva and A.~G.~S.~Landulfo, 
Classical and quantum reconciliation of electromagnetic radiation: 
Vector Unruh modes and zero-Rindler-energy photons, 
Phys. Rev. D \textbf{106}, 065002 (2022).}

\bibitem{vacalis_2023}
{G.~Vacalis, A.~Higuchi, R.~Bingham, and G.~Gregori, 
The classical Larmor formula through the Unruh effect for uniformly 
accelerated electrons, 
Phys. Rev. D~\textbf{109}, 024044 (2024).}

\bibitem{portales_oliva_2024}
{F.~Portales-Oliva and A.~G.~S.~Landulfo, 
Gravitational waves emitted by a uniformly accelerated mass: the role of the zero-Rindler-energy modes in the classical and quantum descriptions, 
Phys.\ Rev.\ D {\bf 109},  045006 (2024).}

\bibitem{brito_2024}
{J.~P.~B.~Brito, L.~C.~B.~Crispino, and A.~Higuchi, 
Gravitational bremsstrahlung and the Fulling-Davies-Unruh thermal bath, 
Phys.\ Rev.\ D {\bf 109},  064080 (2024).} 

\bibitem{unruh-wald_1984}
{W.~G.~Unruh and R.~M.~Wald, 
What happens when an accelerating observer detects a Rindler particle, 
Phys. Rev. D \textbf{29}, 1047 (1984).}

\bibitem{peskin-schroeder} 
{M.~E.~Peskin and D.~V.~Schroeder, 
\textit{An Introduction to Quantum Field Theory}, (Addison-Wesley, Reading, 1995). }

\bibitem{crispino_2008}
{L.~C.~B.~Crispino, A.~Higuchi, and G.~E.~A.~Matsas, 
The Unruh effect and its applications, 
Rev. Mod. Phys. \textbf{80}, 787 (2008).}

\bibitem{Higuchi:2017gcd}
{A.~Higuchi, S.~Iso, K.~Ueda and K.~Yamamoto,
Entanglement of the vacuum between left, right, future, and past: The origin of entanglement-induced quantum radiation,
Phys. Rev. D \textbf{96}, 083531 (2017).}

\bibitem{Ueda:2021nln}
{K.~Ueda, A.~Higuchi, K.~Yamamoto, A.~Rohim and Y.~Nan,
Entanglement of the vacuum between left, right, future, and past: Dirac spinor in Rindler spaces and Kasner spaces,
Phys. Rev. D \textbf{103}, 125005 (2021).}

\bibitem{fulling_1973}
{S.~A.~Fulling, 
Nonuniqueness of canonical field quantization in Riemannian space-time, Phys. Rev. D \textbf{7}, 2850 (1973).}

\bibitem{NIST_website}
{NIST Digital Library of Mathematical Functions, version 1.1.12, \url{https://dlmf.nist.gov/} (2023).}

\bibitem{higuchi_1993}
{A.~Higuchi and G.~E.~A.~Matsas, 
Fulling-Davies-Unruh effect in classical field theory, 
Phys. Rev. D \textbf{48}, 689 (1993).}

\bibitem{sugiyama_2021}
{Y.~Sugiyama, K.~Yamamoto, and T.~Kobayashi, 
Gravitational waves in Kasner spacetimes and Rindler wedges in Regge-Wheeler gauge: formulation of Unruh effect, 
Phys. Rev. D \textbf{103}, 083503 (2021).}

\end{thebibliography}
\end{document}